
\input harvmac

\def\s{\sqrt}
\def\wt{\widetilde}
\def\h{\hat}
\def\[{\left  [}
\def\]{\right ]}
\def\a{\alpha}
\def\b{\beta}
\def\t{\tilde}
\def\({\left (}
\def\){\right )}
\gdef\journal#1, #2, #3, 19#4#5{
{\sl #1 }{\bf #2}, #3 (19#4#5)}

\lref\gk{D. Gross and I. Klebanov, \journal Nucl. Phys., B344, 475, 1990.}

\lref\mo{B. Sathiapalan, \journal Phys. Rev., D35, 3277, 1987.}

\lref\aw{J. Atick and E. Witten, \journal Nucl. Phys., B310, 291, 1988.}

\lref\ev{E. Verlinde }

\lref\kerrbh{See for example S. Hawking, ''Path-Integral Approach to
Quanum Gravity.'', in,
An Einstein Centenary survey, C. U. P., 1979.}

\lref\nk{N. Kaloper, \journal Phys. Rev., D48, 2598, 1992.}

\lref\daveg{D. Garfinkle, \journal Phys. Rev., D46, 4286, 1992.}

\lref\hw{G. Horowitz and D. Welch, \journal Phys. Rev. Lett., 71, 328, 1993.}

\lref\ky{K. Kikkawa and M. Yamasaki, Phys. Lett., 149B, 357, 1984.}

\lref\gth{This argument is due to Gary Horowitz.}

\lref\btz{M. Banados, C. Teitelboim, and J. Zanelli,
 \journal Phys. Rev. Lett., 69, 1849, 1992;
M. Banados, M. Henneaux, C. Teitelboim, and J. Zanelli,
\journal Phys. Rev., D48, 1506, 1993.}

\lref\dvv{R. Dijkgraaf, E. Verlinde, and H. Verlinde,
 \journal Nucl. Phys., B371, 269, 1992.}

\lref\witten{ E. Witten, \journal Phys. Rev., D44, 314, 1991.}

\lref\einstein{In dimensions other than two
the Einstein metric is obtained by conformally rescaling
the string metric by a factor such that the Ricci scalar term in the
action is of the usual form, it is given by
$g_{\a \b}^E = e^{-4\phi /(D - 2)}g_{\a \b}.$}

\lref\hh{ J. Horne and G. Horowitz, \journal Nucl. Phys., B368, 444, 1992.}

\lref\tb{ T. Buscher, \journal Phys. Lett., B201, 466, 1988 ;
\journal Phys. Lett., B194, 59, 1987.}

\lref\rv{ M. Ro\v cek and E. Verlinde, \journal Nucl. Phys., B373, 630, 1992.}

\lref\ghas{G. Horowitz and A. Strominger, Nucl. Phys., B360, 197, 1991.}

\lref\hhs{J. Horne, G. Horowitz, and A. Steif, \journal Phys. Rev. Lett., 68,
568,
1992.}

\lref\noot{This is a slightly different usage from \hhs , here the integrals
over dimensions with translational symmetries are also being taken.}

\Title{\vbox{\baselineskip12pt\hbox{UCSBTH-94-15}
\hbox{hep-th/9405070}}}
{\vbox{\centerline {Timelike Duality}}}

\centerline{\sl Dean L. Welch}
\centerline{\sl Department of Physics}
\centerline{\sl University of California}
\centerline{\sl Santa Barbara, CA 93106-9530}
\centerline{\sl dean@cosmic.ucsb.edu}

\bigskip
\centerline{\bf Abstract}

Several stationary solutions of the low energy string equations are dualized
with respect to their timelike symmetry.
Many of the duals have simple physical interpretations. Those of the
nonextremal three
dimensional black hole and black string are negative mass black strings. The
extremal cases of these, and extremal higher dimensional black strings also,
give negative energy plane fronted waves.
In fact, all of the duals of positive mass solutions that will
be considered here
have nonpositive energies, but an argument is given which suggests that this
is not true in general.

\newsec{Introduction}

One of the interesting properties of string theory
is that it has a symmetry called duality.
The simplest example of duality comes from considering a
flat spacetime with one dimension
compactified to form a circle of radius $R.$ For a theory
based on strings this solution is
equivalent to one where the compactified direction has radius $1/R$ \ky .
However, duality is much more general than this. There exists a more
general duality transformation that maps any solution of the low energy
equations that
has a symmetry to another solution of the low energy equations \tb .
In general these solutions will have very different geometries.
If this symmetry is spacelike and has compact orbits
then both solutions correspond to the
same conformal field theory and
so are equivalent as string theory solutions \rv .

Although duals are usually taken with respect to spacelike symmetries, when a
spacetime is stationary one can take a dual with
respect to its timelike symmetry.
As is the case for a spacelike symmetry, dualizing with respect to a timelike
symmetry maps one solution of the low energy string equations to another.
It is currently not known when these solutions are equivalent as string theory
solutions.
The two dimensional black hole \witten\
\dvv\ has been dualized with respect
to its timelike symmetry and this provides an example of when they are
equivalent (another argument that this will be true in more general
circumstances will be presented shortly).
This duality transformation
maps a solution with positive mass and
a horizon to one with a negative mass naked singularity. It has been
shown that these dual solutions arise from different gaugings of the
same WZW model, therefore they correspond to the same conformal field theory
and are equivalent as string theory solutions. In fact the two dimensional
black hole is its own dual in the sense that this duality transformation
just exchanges parts of the Penrose diagram, the part behind the singularity
is exchanged with the asymptotically flat part.

It is well known that when we dualize a solution with respect to a symmetry
that has a fixed point the dual solution will have a singular geometry.
This is most easily seen
by dualizing flat spacetime written in cylindrical coordinates \rv . If we take
the metric to be $ds^2=-dt^2+dr^2+r^2d\theta ^2,$ dualizing with respect to
$\theta $
gives $\wt{ds}^2 = -dt^2 +dr^2 + r^{-2}d\theta ^2$ (the tilde symbol will be
used to denote dual fields). The dual metric is singular at the fixed point
of $\theta , r=0.$ However, it is nonsingular as a solution to string theory
\rv\ ,
as it must be since it is the dual of a nonsingular solution.

The symmetry that we dualizing with respect to need not have a fixed point
to get such a singularity,
it is sufficient for it to become null.
If $t$ is the timelike coordinate at infinity then we know that $g_{tt} =0$
at, or outside of, an event horizon.
The dual solution will be singular where this happens.
If a black hole has nonzero rotation then
$g_{tt}$ will be zero on the boundary of the ergosphere, which is outside of
the horizon.

The low energy string action is
\eqn\action{ S= \int d^n x \s {-g}\ e^{-2\phi } \[ R +
4(\nabla \phi )^2 - {1 \over 12} H_{\a \b \gamma } H^{\a \b \gamma }
+a \] }
Where, $a$ is a centeral charge term,
$R$ is the Ricci scalar constructed from the spacetime metric,
$H_{\a \b \gamma }$ is a totally antisymmetric three tensor, and
$\phi $ is a scalar field called the dilaton. The equations of motion
that follow from this action are
\eqna\fdeq
$$\eqalignno{&R_{\a \b } + 2\nabla _{\a }\nabla _{\b } \phi
     -{1\over 4} H_{\a \sigma \gamma } {H_\b }^{\sigma \gamma } = 0
       & \fdeq a \cr
 &\nabla ^\a (e^{-2\phi } H_{\a \b \gamma }) = 0   &\fdeq b \cr
 &4\nabla ^2\phi -4(\nabla \phi )^2 + a + R - {1\over 12} H^2 = 0 &\fdeq c
\cr}$$

Just applying the
usual duality transformation \tb\
immediately presents a problem with the transformation of the dilaton.
Simply applying the usual
transformation gives $ \t \phi = \phi -1/2\ln g_{tt},$ but $ g_{tt}< 0 $
outside the horizon. To make sense of this we can
Euclideanize the metric by taking
$t \rightarrow i \tau $ and then dualize with respect to the Euclidean time,
$ \tau $, giving a dual dilaton
of $ \t  \phi =\phi -1/2\ln g_{\tau \tau} =\phi - 1/2\ln ( -g_{t t}). $
Finally, after dualizing we can continue back to a Lorentzian metric.
The fact that the off diagonal
terms, $ g_{t j} $  and  $ B_{t j} $ (as usual $i,j,k,...$ are
indicies other than $t$ ),
temporarily become complex presents no problem, since they become real when we
continue back.
We can either just leave them complex while we dualize or we can continue
other parameters to imaginary values to keep them real.

Applying the above procedure of analytically continuing, dualizing, and
then continuing back we find that
given a solution of the low energy string equations that is independent of
time,
$( g_{\a \b }, B_{\a \b }, \phi )$, then
($\t g_{\a \b }, \t B_{\a \b }, \t \phi $)
is also a solution where \tb

\eqn\dual{\eqalign{
 \t g_{tt} & = 1/g_{tt}, \qquad \quad \t g_{ti} = -B_{ti}/ g_{tt} \cr
  \t g_{ij} & = g_{ij} - (g_{ti}g_{tj} - B_{ti}B_{tj})/g_{tt}  \cr
   \t B_{ti} & = -g_{ti}/g_{tt}, \qquad
         \t B_{ij} = B_{ij} - 2 g_{t[i} B_{j]t}/g_{tt} \cr
     \t \phi & = \phi - {1\over 2} \ln (-g_{tt})\cr }}
Notice that the $tj$ components of the dual metric and
antisymmetric tensor have signs that are
different from what one might have expected (this can be undone by changing
the definitions of the signs of the antisymmetric tensor potential).

It was proven by Ro\v cek and Verlinde that when a spacelike symmetry that we
are
dualizing with respect to has compact orbits then the original solution
and its dual correspond to the same conformal field theory \rv .
The solutions that we are dualizing have horizons. When a solution has a
horizon its Euclidean time must be identified with some finite period
to avoid a canonical singularity. Recall that we obtained our duality
transformation by continuing to Euclidean time and then continuing back
after making the usual duality transformation. The requirement of having
a periodic Euclidean time, together with the results of Ro\v cek and
Verlinde, tell us that the two Euclidean solutions correspond to the
same conformal field theory. This suggests that their Lorentzian signature
counterparts also correspond to the same conformal field theory. This
provides a second, more general, argument that they do. Some evidence
for this will be presented later.

In this work some stationary solutions to the low energy string equations that
have
horizons will be dualized with respect to their timelike symmetry.
The solutions we will consider are: the three dimensional black hole \btz\ \hw\
\nk\ ,
the three dimensional black string \hh\ \hhs\ ,
and black strings in five or more dimensions \hhs\ (which effectively
includes black holes in
four or more dimensions). For all these solutions, except the
three dimensional black hole with
nonzero rotation, $g_{tt}=0$ on the horizons, as stated before this
tells us that the duals
have naked singularities
there. The dual of the rotating three dimensional black hole also has a naked
singularity, but it is where the boundary
of the ergosphere of the original solution was. The tangent to the horizon
of a three dimensional black hole is a Killing vector that is null on the
horizon. One might think it is more natural to dualize the three dimensional
black hole with respect to this symmetry so that the dual solution is singular
on what was the horizon of the original solution. This dual of the three
dimensional black hole seems more similiar to the other dual solutions and will
also be taken.

If in addition we consider spacetimes that are asymptotically flat and that
have a
spacelike translational symmetry in addition to their timelike symmetry,
then we can establish two general results.
It has already been proven \hhs\ that if we dualize an asymptotically flat
spacetime with respect to the spacelike translational symmetry then the
momentum along the symmetry direction per
length, called $P_x $ here, will be
interchanged with the charge of the $H_{\a \b \gamma }$ per length, denoted
$Q.$ In other words we have $\t P_x = Q$ and $\t Q = P_x .$ When we take
one of the these spacetimes that is stationary and dualize with respect to
the time we get a similiar result that differs only by a minus sign, we
get $\t P_x=-Q$ and $\t Q=-P_x .$

In all the cases we will consider the dual solutions of positive mass solutions
will have
nonpositive energy.
Some of the dual solutions are plane fronted waves, these have zero
{\it rest\/} mass and travel at the speed of light.
In these cases one would only talk about the energy of
the solution.
Only the dual of the usual four dimensional black hole, or equivalently
the of five dimensional black string, had zero mass. The other
duals had masses that were negative. This
suggests a general result that dualizing a positive mass solution
with respect to a time gives
a solution that has nonpositive energy. It is not known if this is true,
although
an argument is presented that suggests it is not.
A weaker result concerning the behavior of the energy of the dual solutions
will be established.

\newsec{The Dual Solutions}

\subsec{The Three Dimensional Black String}

The first solution we will dualize is the three dimensional black string \hh .
It is given by
\eqn\bs{\eqalign{ds^2=-\( 1-{M\over r}\) dt^2+\( 1-{Q^2 \over Mr }\) dx^2
+ &\( 1- {M\over r} \) ^{-1}\( 1-{Q^2\over Mr}\) ^{-1}{l^2 \, dr^2\over 4r^2}
\cr
\phi = -{1\over 2} \ln r l \>, \quad & \quad
B_{x t} = {Q \over r}}}
Where $M$ is the mass per length of the string, $Q$ is the charge of the
$H_{\a \b \gamma }$ field per length of the string, and $l$ is related to the
central charge. In addition to the symmetry of translating along $t$ there is
also a spatial translational symmetry in the $x$ direction. This solution
has an event horizon at $r=M$ (for  $Q^2 \le M^2$)
and if $Q \ne 0$ there is also an inner horizon at $r=Q^2/M.$ In the
extremal case, $Q^2 = M^2$, there is only the event horizon. One way to obtain
this solution is to take the product of the two dimensional black hole \witten\
and a translation, this gives us the uncharged three dimensional black
string. To add charge we can boost this solution and then dualize on $x$ \hhs .

Applying \dual\ to this gives
\eqn\bsdual{\eqalign{\widetilde{ds}^2 = \( 1 - {M\over r}\) ^{-1} &\[ -dt^2
- 2{Q\over r}dxdt + \( 1 - {M\over r}\( 1 + {Q^2\over {M^2}}\) \)dx^2 \] \cr
&+ \( 1- {M\over r} \) ^{-1}\( 1 - {Q^2\over Mr}\) ^{-1}{l^2 \,
dr^2\over 4r^2} \cr
\t \phi &= -{1\over 2} \ln l\( r - M\) \>, \quad  \quad
\t B_{\a \b} = 0 }}
Although this may not look familiar it is just a boosted chargeless negative
mass black string. As we will see in the next section the fact that it is
boosted is a manifestation
of the effect dualizing has on the charge and momentum of an
asymptotically flat solution
that also has a spacelike symmetry. What was $H_{\a \b \gamma }$ field
charge in the original solution is now momentum along $x.$
To see that this is a chargeless negative mass black string
we can boost,
and shift $r$, by the coordinate transformation
\eqn\trans{\eqalign{t= \( 1 - {Q^2\over M^2}\) ^{-1/2} \( \h t -
{Q\over M} \h x\) ,
\quad x = \( 1 - {Q^2\over M^2}\) ^{-1/2} \( \h x -
{Q\over M} \h t\), \quad \h r = r - M.}}
In these coordinates the solution is
\eqn\negmass{\eqalign{\widetilde{ds}^2 = - \[ 1 + {M\over \h r}\( 1 -
{Q^2\over M^2}\) \] d\h t ^2 &+ d\h x ^2 + \[ 1 + {M\over {\h r}}
\( 1 - {Q^2\over M^2}\) \] ^{-1} {l^2 d\h r ^2 \over 4 \h r ^2} \cr
\t \phi = -{1\over 2} \ln \h r l \>, \quad & \quad
\t B_{\a \b} = 0 }}
This is an uncharged black string with mass per length $-M(1- Q^2/M^2).$

We can also anticipate this result from conformal field theory
\gth . We know that
the three dimensional black string can be obtained by gauging an axial symmetry
of a WZW model \hh . Suppose we have a WZW model with a symmetry group, $G,$
and we want an axial gauging of some Abelian
subgroup of $G,$ say $H.$ Then the action
of this on an element of $G,$ say $g,$ is $g\rightarrow hgh ,$ where $h$
is an element of $H.$ For
$h$ close to the identity element we get the infinitesimal version of
this, $\delta g = \epsilon g + g \epsilon .$
The action for a vector gauging is $g\rightarrow hgh^{-1},$ the infinitesimal
version of which is $\delta g = \epsilon g - g \epsilon .$

If we take a WZW model with $G=SL(2,R)$ and axially gauge a
certain one dimensional subgroup of $G$ we will get a two dimensional
black hole \witten\ \dvv\ \hh .
If we again take $G=SL(2,R),$ but now vector gauge this subgroup we will get
the dual (we are dualizing with respect to
time since it is the only symmetry for these black holes)
of the two dimensional black hole. This dual has a negative
mass, however the two dimensional black hole is its own dual in the
sense that what we have done by dualizing this solution amounts
to exchanging parts of the Penrose diagram.
The exchange is between the asymptotically flat part
that has the horizon and the part behind the singularity \dvv\ \witten .

Adding one free boson $x$ to the action is equivalent to taking
$G=SL(2,R)\times R.$ If we do this and take the axial gauging of the
same subgroup as in the two dimensional black hole case, and also
allowing the translations to be gauged, we will get the three
dimensional black string \hh . Suppose that we now take the same group,
$G=SL(2,R)\times R,$ and do a vector gauging of the same subgroup.
Since we now have $\delta g = \epsilon g - g \epsilon $ we expect the
$R$ to be unaffected by the gauging because it is Abelian. We
will then have a product of a vecor gauged $SL(2,R)$ and $R$, which is
an uncharged negative mass three dimensional black string.

In addition to showing us that dualizing this solution gives the product
of a negative mass two dimensional black hole and $R,$ it also gives
us the origin of the negative mass.
Following the example of dualizing the two dimensional black hole,
we can interpret this duality as mapping the exterior part of the
black string to the product of $R$ with the
part of the two dimensional black hole that is
behind the horizon.
It also shows us that these solutions correspond to the
same conformal field theory since they are just different
gaugings of the same WZW model.

The extremal black string solution is \bs\ with $Q=M$, the dual
of this is \bsdual\ with $Q=M$. This metric can no longer be
diagonalized, but if we make the coordinate transformation $u=t+x,
v=t-x,$ and $R={l\over 2} \ln (r-M),$ then the solution is in the form
\eqn\exd{\eqalign{\wt{ds}&= -dudv-Me^{-2R/l} du^2 + dR^2 \cr
\t \phi &= -{R\over l} - {1\over 2}\ln l \>, \quad  \quad
\t B_{\a \b} =0 }}
This is just a plane fronted wane in the precence of a linear dilaton
\hhs . Using the mass formula from \hhs\
(for the specific form of the mass formula see section 3)
, which really gives us the energy
of the solution, shows that this has a negative energy.

It is interesting to compare these results with what we get by taking a three
dimensional black string and dualizing it with respect to some
spacelike symmetry. For the nonextremal case dualizing it with respect to
$x$ gives an uncharged boosted black string, with velocity depending on
Q \hhs . We get the three dimensional black hole (to be discussed below)
by dualizing with respect
to $\partial /\partial t + \partial /\partial x $ \hw . This symmetry is
spacelike
everywhere outside of the horizon, but it becomes null at infinity.
If we dualize the extremal three dimensional black string with respect
to $x$ we get a positive energy plane fronted wave. To summarize, for the
three dimensional black string dualizing with respect to $t$ gives
negative energy versions of the
same solutions as dualizing with respect to $x$ does.

\subsec{The Three Dimensional Black Hole}

The three dimensional black hole solution is \btz\ \hw\
\eqn\bh{\eqalign{ds^2 = -{1\over l^2}\( r^2  - r_+^2 - r_-^2 \)
 dt^2 & -
{2r_+r_-\over l}dtd\theta + r^2d\theta ^2 + {l^2r^2dr^2\over \( r^2 - r_+^2\)
\( r^2 - r_-^2\) } \cr
\phi = 0, \quad & \quad  B_{t \theta } = -{r^2\over l}}}
The mass of this black hole is $M_{bh}=(r_+^2 + r_-^2)/l^2$ and its angular
momentum
is $J=2r_+r_-/l $, without loss of generality $J\ge 0 $ is being taken.

This metric is asymptotically anti-de Sitter.
It has two horizons, an event horizon located at
$r=r_+$ and an inner horizon at $r=r_-.$ This solution has many of
the features one usually associates with a black hole, in addition to the
horizons it has an ergosphere and a nonzero Hawking temperature. However,
it also has some unusual features for a black hole, its curvature is
constant, $l$ is related to the curvature by $R_{\a \b} = -(2/l^2 )
g_{\a \b} $, this also tells us it has no curvature singularity (when $J=0$ it
has a Taub-NUT type singularity, if $J\ne 0$ it is singularity free).
Unlike the three dimensional black string this has two distinct
extremal limits, $J^2=l^2M^2 \ne 0$ and $J=M=0$.

The nonextremal solution, $J^2<l^2M^2 ,$
will be considered first. Applying \dual\ yields

\eqn\bhdual{\eqalign{\wt{ds}^2 = {l^2\over r^2 - r_+^2 - r_-^2}&\[ -dt^2 -
2{r^2\over l}dtd\theta - {r^2 \( r_+^2 + r_-^2\) - r_+^2r_-^2 \over l^2}
d\theta ^2 \]  \cr
&+ {l^2r^2dr^2\over \( r^2 - r_+^2\) \( r^2 - r_-^2\) } \cr
\t \phi = -{1\over 2}&\ln {r^2 - r_+^2 - r_-^2 \over l^2} \>, \quad  \quad
\t B_{t \theta } = -{lr_+r_-\over r^2 - r_+^2 - r_-^2}}}
Notice that this becomes singular at $r^2 = r_+^2 + r_-^2,$ this is at what
was the boundary of the ergosphere and occurs before we reach the
horizon of the original solution. We now have not only $\partial /\partial t$
being timelike, but $\partial /\partial \theta $ being timelike as
well. However, the signature of the dual spacetime is the same as the
original, as will be seen shortly.

If we make the coordinate transformation
\eqn\transb{\eqalign{t = {r_+^2\h x - r_-^2\h t\over l\s {r_+^2 - r_-^2}},
\quad  \theta
= {\h t - \h x \over \s {r_+^2 - r_-^2}}, \quad  l\h r = r^2
- r_+^2 - r_-^2}}
Then these fields \bhdual\ become
\eqn\bhdualb{\eqalign{\wt{ds}^2 = -\( 1 + {r_+^2\over l\h r }\) d\h t^2 +
\( 1 + {r_-^2\over l\h r}\) d\h x ^2 &+ \( 1+{r_+^2\over l\h r}\) ^{-1}
\( 1 + {r_-^2\over l\h r}\) ^{-1} {l^2d\h r^2\over 4\h r ^2} \cr
\t \phi = -{1\over 2}\ln {\h r\over l} \>, \quad & \quad
\t B_{\h t \h x }={r_+r_- \over l\h r}}}
Comparing this with \bs\ we see that this is a black string with
$M=-r_+^2/l$ and $Q=-r_+r_-/l$.
We also have time being periodic,
since $\theta $ is periodic, so we must
go to a covering space to avoid closed timelike curves.

Now the first extremal case will be considered, $M=J=0.$ Taking \bhdual\ with
$r_+=r_-=0$ and making the coordinate transformation $u=t, v=2l\theta , R =
l\ln r,$ we have
\eqn\exbhda{\eqalign{\wt{ds} ^2 &= -l^2 e^{-2R/l}du^2 - dudv + dR^2 \cr
\t \phi &= -{R\over l} -\ln l \>, \quad  \quad
\t B_{\a \b } = 0 }}
Which is again a plane fronted wave in the presence of a linear dilaton,
it also has negative energy.

Now consider the other extremal case, $J = Ml \ne 0 .$ This corresponds
to inserting $r_+ =r_- \ne 0$ into \bhdual\ . Doing this and making the
coordinate change, $t = u - vr_+^2/2l^2, \theta = v/2l,$ and
$lR= r^2 - 2r_+^2 $, the metric becomes
\eqn\exbhb{\eqalign{\wt{ds}^2 = -{l\over R}du^2 - \( 1 + {r_+^2\over lR} \)dudv
+ {l^2dR^2\over 4 \( R + r_+^2/l \)^2}}}
This is just a negative energy
plane wave superimposed on a charged negative mass black
string \daveg\ \hw .

With the exception of a three dimensional black hole with nonzero rotation,
$\partial /\partial t $ is tangent to the horizon for the solutions
considered here. Since the horizon is a null surface, the dual with
respect to time will be singular where the horizon of the original solution
is, when $\partial /\partial t$ is tangent to the horizon. For the three
dimensional
black hole $\partial /\partial t$ is null at $r=\s {r_+^2+r_-^2}$, which is the
location of the boundary of its ergosphere. If $J\ne 0$ then this will be
outside of the horizon, which is at $r=r_+.$ To make the dual of the three
dimensional black hole more similiar to the other cases one might want
to dualize it with respect to the tangent of the horizon.

The vector $\partial /\partial \chi = \partial /\partial t + {r_-\over lr_+}
\partial /\partial \theta $ is tangent to the horizon. The metric \bh\ tells
us that $(\partial /\partial \chi )^2 = -(r_+^2+r_-^2)(r^2-r_+^2)/l^2,$ so
$\partial /\partial \chi $ is null at the horizon, as it had to be, and
timelike outside of it. When $J=0$ this reduces to $\partial /\partial t.$
The vector $\partial /\partial \chi $ is a Killing vector so we can
dualize the three dimensional black hole with respect to it. Doing this
gives an uncharged three dimensional black string with mass
per length $-(r_+^2 - r_-^2)/l,$ as before this is negative. Recall that
if the three dimensional black hole had nonzero rotation then dualizing it
with respect to $t$ gave a charged negative mass black string.

Like the three dimensional black string, the three dimensional black hole
has a spacelike Killing vector, $\partial /\partial \theta .$ The results
of dualizing the three dimensional black hole with respect to this
symmetry are known \hw\ \nk . The nonextremal case gives us a positive
mass three dimensional black string, if the black hole has $J\ne 0$ then
the black string will have a nonzero charge. The $M=J=0$ extremal case
gives a positive energy plane fronted wave and the $J=lM \ne 0$
extremal case gives a plane fronted wave superimposed
on an extremally charged black string, both with positive energy.
As in the case of the three dimensional black string, except for having
negative energies, dualizing a three dimensional black hole with respect to
its timelike symmetry gave the same result as dualizing with respect to its
spacelike symmetry.

\subsec{Black Strings In Five Or More Dimensions}

Taking $D$ to be the number of spacetime dimensions, the charged black string
solutions for $D\ge 5$ are given by \hhs\ ,
\eqn\higher{\eqalign{ds^2 = &-{1 - M_o/r^n \over 1 + M_o\sinh ^2 \a /r^n} dt^2
+{dx^2\over 1 + M_o\sinh ^2\a /r^n }  \cr &+
{dr^2\over 1 - M_o/r^n } + r^2d\Omega ^2_{n+1} \cr
\phi = &-{1\over 2}\ln \( 1 + {M_o\sinh ^2\a \over r^n} \) \>, \quad  \quad
B_{t x} = - {M_o\cosh \a \sinh \a \over r^n + M_o\sinh \a } }}
The mass per length of this black string is $M= M_o(1 +n/(n + 1) \sinh ^2\a )$,
its charge per length is $Q= n/(n + 1) \sinh \a \cosh \a $,
and $n$ is related to
the spacetime dimension by $n = D - 4 . $ Like the three dimensional black
string this solution has translational symmetry and can
be constructed in a manner
similiar to that of the three dimensional case.
These black strings can be constructed by taking the product of
a $D - 1 $ dimensional Schwarzschild black hole with $R$, then boosting
with a velocity parameter $\a $, and dualizing
with respect to $x$ \hhs . A $D$ dimensional black string with $\a = 0$ is the
product of a $D-1$ dimensional Schwarzschild black hole and $R.$
This tells us that the results we get by dualizing
the black strings in $D$ dimensions can be used to get the duals of the
$D - 1$ black holes by taking $\a = 0.$

Using \dual\ to dualize \higher\ we find that
\eqn\higherdual{\eqalign{\wt{ds}^2 &= -{r^n + M_o\sinh ^2\a \over r^n - M_o
}dt^2
- {2M_o \sinh \a \cosh \a \over r^n - M_o}dxdt
\cr &+ {r^{2n}-r^nM_o - M_o^2\sinh ^2\a \cosh ^2\a \over \( r^n - M_o\) \(
r^n + M_o\sinh ^2\a \) }dx^2 + {r^n \over r^n - M_o }dr^2 + r^2 d\Omega ^2
_{n+1}\cr
\t \phi &= -{1\over 2}\ln \( 1 - {M_o \over r^n }\) \>, \quad  \quad
\t B_{\a \b } = 0 }}
As expected from the general result, that will be established
later, of charge and momentum being interchanged
(with a negative sign) we have an uncharged boosted product being the result of
dualizing a charged solution that has zero momentum.

If we make the coordinate transformation $t = t^\prime \cosh \a -
x^\prime \sinh \a $ and $x = x^\prime \cosh \a  - t^\prime \sinh \a $
the metric becomes
\eqn\hdbsdualb{\eqalign{\wt{ds}^2 = {r^n\over r^n - M_o}dt^{\prime 2}
+dx^ {\prime 2} + {r^n\over r^n - M_o}dr^2 + r^2 d\Omega ^2 _{n+1}}}
This is the product of the dual of a $D-1$ dimensional black hole
and a translation. It has mass per length $-(D-5)/(D-3)M_o.$

We can get the mass of the dual of a black hole in four or more dimensions
from this by taking $D \rightarrow D + 1 $ and $\a = 0$. This gives us
$\t M=- (D-4)/(D-2)M_o,$ where $M_o$ is the mass of the original black hole.
It is interesting to note that the dual of the usual four dimensional
Schwarzschild black hole has zero mass.

The extremal $D \ge 5$ dimensional black string is obtained by taking $M_o
\rightarrow 0$ and $\a \rightarrow \infty $ in such a way that
$M_o\sinh ^2\a $ approaches a constant, clearly this gives $M=Q$.
If we do this in \higherdual\ and make the coordinate transformation
$u = t + x $ and $v = t - x $
we will get
\eqn\exhigher{\eqalign{\wt{ds} ^2 = -dudv - {M_o\sinh ^2\a \over r^n}du^2 +
dr^2 + r^2d\Omega ^2_{n + 1}}}
This is just a plane fronted wave in $D \ge 5$ dimensions \hhs . This solution
also has negative energy.

Finally we can compare these duals with the results of dualizing the higher
dimensional black string \higher\ with respect to $x$ \hhs .
Dualizing the nonextremal case with respect to $x$ gives an uncharged boosted
positive mass black string that is boosted with a velocity parameter, $\a ,$
implicitly
related to the charge per length of the original black string by
$Q = n/(n + 1)\sinh \a \cosh \a .$ For the extremal case dualizing with respect
to $x$ gives a positive energy plane fronted wave, recall that dualizing the
extremal solution with respect to $t$ gives a negative energy plane fronted
wave.

\newsec{General Results}

We can get two general results if we
consider solutions that are  asymptotically flat in the sense that \hhs\
asymptotically we have $g_{\a \b } \rightarrow \eta _{\a \b } +
\gamma _{\a \b },
\phi \rightarrow \phi _o + \chi , B_{\a \b } \rightarrow B_{\a \b }$
such that $\gamma _{\a \b }, B_{\a \b }, $ and $\chi $ all approach zero.
We want to allow solutions with translational symmetry to be asymptotically
flat, so large distances will refer to large transverse distance. The
proper distance from the source will be denoted by $r.$ If we let
$N = D$ for black holes
and $N = D - 1 $ for black strings, then for $N\ge 4 $ we require these
fields to approach zero at least as fast as $r^{-(3-N)}.$
If $D=3$ then the requirement for asymptotic flatness is that they approach
zero at least as fast as $e^{-ar}$, with $a$ being some constant. Only the
black
string solution is relevant here, there do not appear to be asymptotically
flat black holes in three dimensions or asymptotically flat black strings
in four dimensions.
We also require $\phi _o=-ar/2 +b,$ where $b$ is also a constant. For
$D\ge 4$ $a$ must be zero.

If we take such a solution and apply a duality transformation with
respect to a vector that has a norm that approaches
a nonzero constant asymptotically, then the dual will also be asymptotically
flat. The three dimensional black string is of this form and the three
dimensional
black hole is not. Even though these solutions are duals of each other \hw\
\nk\
(see also section 2.2), there
is no conflict because to go from the
asymptotically flat solution to the one that is not, one must dualize with
respect to a symmetry that is asymptotically null.
Given such an asymptotically flat solution and applying \dual\ one finds that
to first order the asymptotic parts of the fields transform as
\eqn\asydual{\eqalign
{\t \gamma _{t t} &= - \gamma _{t t},\quad  \t \gamma _{i j} =
\gamma _{i j},\quad  \t \gamma _{t j} = B_{t j}, \cr
\t B_{t j} &= \gamma _{t j}, \quad  \t B_{i j} = B_{i j}, \quad  \t \chi = \chi
+
{1\over 2} \gamma _{t t} }}

To get our first general result suppose that
our spacetime has a spacelike translational symmetry in
addition to the timelike one, such as the black string cases. As in these cases
call this the $x$ direction. We can calculate the momentum per length
in the $x$ direction and the charge of $H_{\a \b \gamma }$ per length by using
\hhs\
\eqn\momchar{\eqalign{P_x = -C\oint e^{-2\phi } \( \partial _r \gamma _{tx}
- \partial _t \gamma _{x r}\) \cr
Q= C \oint e^{-2\phi}\( \partial _r B_{t x} + \partial _t B_{xr}\) }}
where $C$ is a dimension dependent constant and the integrals are over
closed surfaces at large $r.$
It has been shown \hhs\ that when we dualize
with respect to $x$ we have $\t P_x = Q $
and $\t Q = P_x .$ If we want to dualize
with respect to $t$ we need $g_{\a \b }, B_{\a \b },$
and $\phi $ all to be independent of $t.$ This tells us that the second terms
of
$P_x $ and $Q$ are zero. Therefore, when we dualize
with respect to $t$ \asydual\ shows us that
$\t P_x = - Q$ and $\t Q = - P_x .$ Up to a sign this is the same result
as dualizing with respect to $x.$

The formula for the mass of a solution is \hhs\
(more accurately this is a formula for
the time component of the momentum, so it is really the energy)
\eqn\mass{\eqalign{M=C \oint e^{-2\phi } \[ \nabla ^i \gamma_{i j } -
\partial _j \( \gamma - 4 \chi \) -2\gamma _{i j}\partial ^i\phi _o\] dS^j}}
where $\gamma $ is the trace of $\gamma _{i j},$
$\nabla _i$ is the covariant derivative associated with
the coordinates one chooses at infinity (for example $\nabla _i = \partial _i$
if we have a Cartesian coordinate system at infinity),
and the integral is over a surface at large $r$---since
we are allowing translational symmetries asymptotic
flatness refers to large transverse distances
from the object \noot . If we dualize with respect to a spacelike
symmetry, say $x$, it has been shown that this is invariant \hhs . The
reason is that the change in $\chi $ under duality is cancelled
by that in $\gamma _{x x}.$ When we dualize with respect to
time we now do not change
the spatial metric to first order, but the dilaton still changes. In fact
one gets
\eqn\massdual{\eqalign{\t M = C \oint e^{-2\phi }\[ \nabla ^i \gamma _{ij}
- \partial _j\( \gamma - 4\chi \) - 2\gamma _{i j}\partial ^i \phi _o +
2\partial _j \gamma _{t t}\] dS^j }}
for the mass of the dual solution.
It is easily seen that if one has a solution where $\partial _r \gamma _{t t}
< 0, $ then the mass of the dual is less than the mass of the
original. This is the second general result.

The condition $\partial _r \gamma _{tt} < 0$ is basically
that we get signals redshifting as they move away from the source. Intuitively
one would expect positive mass solutions to exihibit this behavior. However,
$\gamma _{tt}$ is the asymptotic part of the string metric and it is really the
behavior of the Einstein metric
\einstein\ that determines the mass.

If it were true that all positive mass solutions had $\partial _r g_{tt} < 0$
at infinity then one could show that the dual with respect to time of a
positive mass solution has nonpositive mass. The reasoning is as follows.
Suppose it were true, then if we were dualizing a positive mass solution
we would know that $\partial _r\gamma _{tt} < 0$, using \massdual\ we
would then know that $\t M < M.$
Suppose that we also had $\t M > 0,$ then if we were to dualize with respect to
$t$ again we would then have $\t {\t M} < \t M,$ by using \massdual\ again.
However we know that dualizing with respect to the same symmetry twice gives
one the solution one started with, therefore we know that $\t {\t M} = M.$
So by assumming that $\t M >0 $ we have arrived at a contradiction,
therefore $\t M$ must be nonpositive. Thus, if it were true
that positive mass solutions always have
$\partial _r \gamma _{tt} <0$ then we
would know that a positive mass solution dualized with respect to time would
give a solution with nonpositive mass, or more accurately energy.

Unfortunately its not clear that all positive mass solutions will have
$\partial _r\gamma _{tt} < 0$. Consider the dual of the four dimensional black
hole, it has zero mass and $\partial _r \gamma _{tt} = M/r^2 ,$ where $M$ is
the
mass of the black hole that we dualized, so it can take arbitrary values.
Suppose we add a small amount of positive mass to this solution. If the mass
is small enough its effect on $g_{tt}$ at infinity should be small, but as long
as we add a finite amount of positive mass to this we should have a positive
mass solution. If this is correct then
this would be an example of a spacetime with an asymptotic
blueshifting string metric, $\partial _r \gamma _{tt} > 0$, and a positive
mass, leaving us with no general result.
More than this, we would now have a positive mass solution that yields another
positive mass solution when dualized with respect to time. This follows from
the fact that if $\partial _r\gamma _{tt} > 0$ then \massdual\ shows us
that $\t M > M,$ so $M>0$ then leads to $\t M > 0.$

\newsec{Conclusions}
In this work several stationary solutions to the low energy
string theory equations were considered.
The fact that they have a timelike Killing vector allows one to dualize them
with respect to that vector.
All of the solutions dualized have horizons and since $\partial /\partial t $
becomes null at, or before, the horizon we expect all of these duals to have
naked singularities.
For asymptotically flat solutions that have a spacelike symmetry in addition to
the
timelike one, it was found that dualizing the time exchanged the momentum along
the spatial symmetry with the negative charge of the antisymmetric three
tensor. Except for the change of sign this is the same result one would get
by dualizing the translational symmetry.

All of the examples dualized had positive mass and all of their duals had
nonpositive
energy. This suggests the possibility of the existence of a general
result that this will always be the case.
Not only was no general result found, an apparent counterexample to this was
described.

Several of the duals gave familiar solutions. The duals of the extremally
charged black strings gave negative energy plane fronted waves. The duals
of the extremally rotating three dimensional black holes, the $M=J=0$ case
and the $J=lM\ne 0$ case, gave negative energy plane fronted waves, in the
later
case this was superimposed on a negative energy charged black string.
The dual of the nonextremal three dimensional black hole is a negative mass
black string, if the black hole had nonzero rotation then the black string
will be charged. The dual of the three dimensional black string is an
uncharged negative mass black string. An explanation of this which shows that
these solutions correspond to the same conformal field theory was also given.

The fact that these duality transformations
map positive mass solutions to ones with negative
mass may seem strange because, as stated before, we expect these solutions to
be equivalent as string theory solutions. The resolution to this is that we are
presumably mapping the exterior region of a solution that has positive
mass to a region that has negative mass, but that is only part of a larger
space.
This being a generalization of the effect of
dualizing a two dimensional black hole or a three dimensional black string.

\vskip 1cm

\centerline{Acknowledgments}
I would like to thank Gary Horowitz for many useful discussions and for reading
earlier drafts of this paper. I would also like to thank the Issac Newton
Institute for Mathematical Sciences and Cambridge University for providing
a stimulating atmosphere while part of the work was being done. This work
was supported in part by NSF Grant PHY-9008502.

\listrefs

\end